\documentstyle[11pt,epsfig]{article}
\input{epsf}
\def\be{\begin{eqnarray}}
\def\ee{\end{eqnarray}}
\def\ben{\begin{equation}}
\def\een{\end{equation}}
\def\bi{\bibitem}

\def\prl{Phys. Rev. Lett.}\def\pr{Phys. Rev.}\def\np{Nucl. Phys.}
\def\pl{Phys. Lett.}

\def\roughly#1{\mathrel{\raise.3ex\hbox{$#1$\kern-.75em%
\lower1ex\hbox{$\sim$}}}}

\def\itt{\indent\indent}

\def\sigmab{\mbox{\boldmath $\sigma$}}

\def\Bnabla{\mbox{\boldmath $\nabla$}}

\def\A0{A_0}
\def\bq{\begin{equation}}
\def\eq{\end{equation}}

\renewcommand{\thefootnote}{\fnsymbol{footnote}}
\begin{document}
\begin{titlepage}
\begin{center}
%\hfill {KIAS-P000yy}

 \vskip 2cm {\Large \bf Physics of Dense and Superdense Matter\footnote{
Invited talk given at the First KIAS International Workshop on
Astrophysics: Explosive Phenomena in Compact Astrophysical
Systems, KIAS, Seoul, Korea; 24-27 May 2000.}}
  \vskip 1cm
   {{\large Mannque Rho$^{a,b}$} }
 \vskip 0.2cm

 {\it (a) Service de Physique Th\'eorique, CE Saclay, 91191
Gif-sur-Yvette, France}

{\it (b) School of Physics, Korea Institute for Advanced Study,
Seoul 133-791, Korea}

\end{center}

\vskip 1cm

\centerline{(\today)}
 \vskip 1cm

\centerline{\bf Abstract} \vskip 0.5cm I discuss a few aspects of
dense hadronic matter and superdense QCD matter that are
considered to be relevant to the physics of compact astrophysical
systems. The connection between a ``bottom-up approach" and a
``top-down approach" is made with the help of an effective field
theory strategy. Topics treated are meson condensation going up
from low density and color superconductivity going down from
asymptotic density with the approach to the chiral phase
transition made in terms of Brown-Rho scaling.

\end{titlepage}
\newpage

\renewcommand{\thefootnote}{\arabic{footnote}}
\setcounter{footnote}{0}

\begin{figure}[thb]

\vspace{-1.1cm} \epsfxsize=8cm
\begin{center}
\hspace*{0in} \epsffile{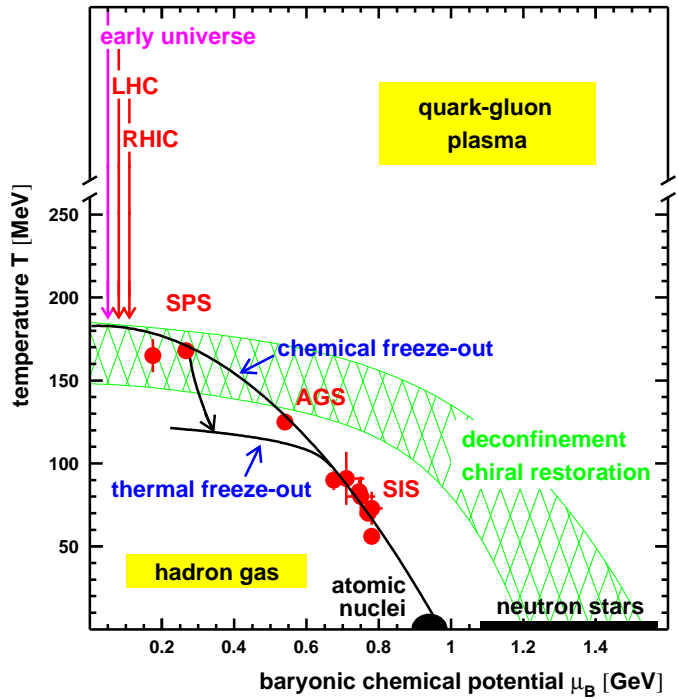}
 %\hspace*{0in} \epsfig{file=phase_new.eps}
\vspace{-2ex}
\end{center}

\vspace{-1.0cm}

\caption{ The conjectured QCD phase diagram. It shows the recent
development including the hadro-chemical freeze-out points
determined from thermal model analyses of heavy-ion collision
data at SIS, AGS and SPS energy. The hatched region represents
the plausible phase boundary delineating two different phases of
QCD indicated by lattice QCD calculations at $\mu_b$=0.}
\label{phase}
\end{figure}

{\section{Introduction}} \itt Super-dense hadronic and/or quark
matter is considered to be relevant for understanding compact
stars in the context of supernova explosions, hypernovae, merging
of compact (neutron and black-hole) stars, gamma-ray bursts etc.
See \cite{shapiro} for a recent review. It is not known at present
precisely which densities are relevant to such systems but the
region of the density involved is presumably somewhere in-between
very low and very high density regimes, the two extreme regimes
being in principle fairly well controlled. To study the physics
of the density regime involved which cannot be systematically
accessed as I will describe below, we have at our disposal
basically two possible theoretical approaches: One, ``bottom-up"
one going up from very low density and the other, ``top-down"
going down from very high (asymptotic) density. The former relies
on laboratory (experimental) data available at low density in
nuclear and hadronic physics and the latter on the theoretical
machinery available at super-high density at which QCD becomes
weak-coupling. We would eventually like to bring the two
approaches together and have them match. But both going up beyond
nuclear matter density and going down from the asymptotic density
to the relevant regime are difficult because of lack of reliable
theoretical tools and of experimental data. In this talk I will
discuss some aspects of both and indicate how the two ``extremes"
could be brought together at a density regime supposedly relevant
to the astrophysical processes of interest. This talk will be
qualitative, intended to give a personal overview of the present
situation. Certain aspects of the topic somewhat lying outside of
the main theme of this meeting are discussed in a separate
note~\cite{migdal-90}.
 \vskip 0.5cm
{\section{Phase Structure}}
 \itt
 In many of the current meetings in nuclear and hadronic physics, one is
shown a canonical diagram depicting the phase structure of
hadsronic matter in terms of temperature (T) vs. chemical
potential $\mu$ or baryon number density $\rho$. In heavy-ion
meetings, the principal focus is put on the phase structure at
high temperature. There, the phase below $T_c\sim 150$ Mev is in
hadronic state and the one above in quark-gluon plasma. There is
strong evidence from lattice gauge calculations that the phase
transition does indeed take place at a given $T_c$ although above
the $T_c$, things are not as simple as the naive form of
quark-gluon plasma that people have thought of before. In any
event, it seems likely that one will be able to follow
temperature properties of the systems with small non-zero density
both theoretically and empirically, the former on lattice and the
latter in the heavy-ion machines that are operating and/or are to
come in the near future~\cite{satz,schukraft}. The most recent
phase diagram taken from Braun-Munzinger's article~\cite{peter}
-- which was also seen in Schukraft's talk -- is shown in
Fig.\ref{phase}~\cite{peter}. It gives a perspective in this
matter. There are many laboratories involved in this endeavor,
such as ``early universe," LHC, RHIC, ..., SIS as indicated in
the figure.

It is a different story with matter at high density. For
technical reasons, it turns out to be extremely difficult to put
dense matter on the lattice except for the academic case of
$N_c=2$ systems, accounting for the paucity of theoretical
guidance as to what happens to hadronic systems at non-zero
density. The knowledge we have up to date is based on various
models.  Models indicate that as density increases toward the
point at which QCD (chiral) phase transition is supposed to take
place, there can be a multitude of phase transitions that may or
may not be in concordance with chiral phase transition. As I
shall discuss below, there are some experimental data around
nuclear matter density but there are no terrestrial data much
beyond. Thus there is need to resort to astrophysical
observations. Neutron stars are the only laboratory available for
this purpose as indicated in Fig.\ref{phase}. The aim of this
note is to discuss how to proceed along the density axis of the
figure.

 \vskip 0.5cm
 {\section{Probing Dense Matter}}
 \itt
 Matter at high temperature which may be relevant for the
early Universe can be -- and is being -- probed in heavy-ion
collisions at relativistic energy. Furthermore QCD at high
temperature can be simulated on lattice which will eventually
give a solid theoretical information. Now what about probing
matter at high density? Up to nuclear matter density, ample
information is available from nuclear systems. Thus combined with
reasonable models, one could attempt to go beyond nuclear matter
by extrapolating what we learn at nuclear matter density. This
constitutes the ``bottom-up" approach. At low density, hadrons
are the proper variables in QCD and the ``bottom-up" approach
consists of doing systematic calculations in the strong-coupling
regime aided by experimental data. However one cannot push this
strategy too far: this scheme must break down at some density.

At asymptotic density, the theory is in principle well-defined
since QCD is weak-coupling there and perturbation theory is
well-defined. One may therefore start with the weak-coupling QCD
and adopt the ``top-down" approach. As we go down in density,
however, the coupling increases, so at some point the perturbative
approach must also break down. The ultimate goal would then be to
have the two approaches come together at some intermediate
density. The difficulty is that the gap between the two
break-down points is not narrow and it is in this gap that
interesting astrophysical phenomena seem to take place.
Furthermore, the region in question cannot be probed in
terrestrial laboratory experiments. Thus no reliable predictions
can be made. We believe that the compact stars such as neutron
stars are the only systems from which we can gain empirical
information for the regime concerned and since no lattice QCD is
available at finite density at least in the near future, the
compact stars seem to be the only direct sources for dense matter
physics.
 \vskip 0.5cm
 \section{Bottom-Up Approach}
{\subsection{Pions Do Not Condense}}
 \itt
 Beyond the normal matter density, the first phase change that
is expected to occur is meson condensation. The lightest meson is
the pion so one would think that pions would condense
first~\cite{migdal}. What happens can be understood in a simple
and caricatured way as follows.

Pions may condense in S or P waves. The S-wave condensation can
occur due to a change in the pion mass in medium. This is a
simple reflection of the fact that the mass term in scalar field
theory is a ``relevant"  term. Since the pion mass in nature is
$\sim 140$ MeV which is small compared with the chiral symmetry
scale, $\Lambda_\chi\sim 1$ GeV, a small attraction in the scalar
direction could push the mass down to zero, so one would think
that it would be easy to trigger the phase transition with the
order parameter $\langle \pi^a\rangle\neq 0$. Baryon matter
density $\rho\sim \langle B^\dagger B\rangle$ where $B$ is the
baryon field could induce such a change which will be an S wave.
In nature this phase transition does not seem to occur mainly
because there is chiral symmetry protection. Although the pion is
massive, its mass is small because chiral symmetry is broken
explicitly by a small amount, i.e., the tiny quark mass which
arises from the symmetry breaking at the electro-weak scale. Thus
in some sense chiral symmetry is still operative and hence
protects the pions from undergoing a phase change in the S-wave
channel.

On the other hand, the P-wave condensation can occur due to the
P-wave attraction that pions feel in the presence of nucleons.
The interaction dictated by chiral symmetry is of the derivative
type $\sim \frac{g_A}{f_\pi}\overline{N}^\dagger (x)
\tau^a\sigmab N(x)\cdot\Bnabla \pi^a (x)$ where $g_A$ is the
axial coupling constant and $f_\pi$ is the pion decay constant,
both of which are associated with the weak interaction. This
P-wave interaction with the nucleon is strong as evidenced by the
huge resonance $\Delta (1230)$ associated with the strong $\pi N$
interaction. The attraction is greater the greater the coupling
$g_A$. The P-wave attraction which increases with density in
nuclear medium can generate instability against P-wave pion
condensation signaled by a space-dependent order parameter
$\langle \pi^a (x)\rangle$. What happens as density increases can
be analyzed with a Lagrangian that preserves chiral symmetry
structure of dense matter~\cite{baym}. The crucial element in this
formulation of pion condensation is that the phase transition is
extremely sensitive to how $g_A$ -- which is associated with the
space component of the axial current -- behaves in nuclear medium.
It turns out that the critical density goes like
$\rho_c^{\pi}\propto (g_A^\star -1)^{-1}$. One can therefore get
a qualitative idea as to how the P-wave condensation occurs by
looking at how the effective coupling $g_A^\star$ behaves in
nuclear medium.

There has been much controversy on the effective $g_A$ in nuclei
and nuclear matter and its relevance to chiral phase structure.
Experiments in nuclei exciting giant Gamow-Teller resonances have
been interpreted in terms of this effective Gamow-Teller coupling.
Whatever may be the valid interpretation, the outcome is that the
{\it effective} $g_A$ (that is, $g_A^\star$ which is density
dependent)\footnote{There has been much debate as to whether the
quenching of $g_A$ in nuclei and nuclear matter as evidenced in
Gamow-Teller transitions is {\it fundamental} or some
garden-variety nuclear phenomenon. I believe this is a non-issue.
There is no question that taken as an {\it effective} parameter
in a precisely defined sense, the $g_A$ is quenched in nuclei.
Whether it can be described by many-body effects (e.g., ``core
polarizations" due to tensor forces) or $\Delta$-hole effects or
even partial chiral restoration is mostly semantic -- they are of
the same physics -- and is not relevant for the issue. What is
relevant is that the {\it effective} Gamow-Teller matrix element
can be described in dense medium by a quenched $g_A^\star$ in the
chiral Lagrangian and that it goes to the value of 1 as density
increases.} tends to the limit
 \be
 g_A^\star\rightarrow 1\ \ \ {\rm as} \ \ \rho\rightarrow \rho_0.
 \ee
 This argument does not pinpoint the critical density but one can
 say that $\rho_c^\pi$ is pushed beyond the regime where the
 theoretical framework makes sense.
  \vskip 0.5cm
\subsection{Kaon Condensation}
 \itt
 As we will hear in the next talk by Chang-Hwan Lee, kaon
 condensation can occur in, and make an important impact on,
 neutron-star formation. I will give a very simplified argument
for how and why kaons will condense in dense neutron-star matter.
More rigorous discussion is given by Chang-Hwan Lee in
 \cite{chlPR}.

\begin{figure} \centerline{\epsfig{file=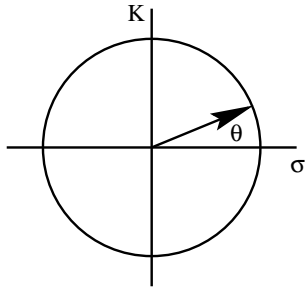,width=4cm}}
\caption{\small Projection onto the $\sigma, K$ plane. The
angular variable $\theta$ on the ``chiral circle" represents
fluctuation toward kaon mean field. Kaon condensation corresponds
to ``rotating" into the $K$ direction (slightly) away from the
chiral circle.}\label{Ksigma}
\end{figure}

 In a nut-shell, a kaon, particularly the negative charged $K^-$
 which contains an anti-strange quark $\bar{s}$, can condense in
 neutron-star matter precisely because the strange quark is
 massive at the scale of low-energy physics we are concerned with.
 Since the strange quark has a current mass of $\sim 150$
 MeV~\footnote{The quark mass is not renormalization-group
 invariant and so has to be defined with a given scale. Even at a
 given scale, it is not well-determined. Various methods of analysis
give the range $100\sim 200$ MeV for the strange current-quark
mass. This uncertainty makes the estimate of the critical density
somewhat uncertain.}, chiral symmetry is explicitly broken in the
strange sector and the low-energy theorems associated with Goldstone
bosons sometimes fail badly. What the matter density does is to
``rotate" the explicitly broken symmetry in such as way that the
kaon mass gets lighter to partially restore the explicitly broken
symmetry. Suppose that all Goldstone boson masses were zero. Then
the same chiral symmetry protection argument that applies to
pions will apply to kaons as well so that baryon number density
cannot do anything to the energy density of the system. If on the
other hand all Godlstone bosons were massive and degenerate, the
rotation around the chiral circle (see Fig.\ref{Ksigma}) would
not affect the energy density and so nothing would happen.
Indeed, things happen precisely because the potential is tilted
due to the symmetry breaking and the baryon density does just the
un-tilting to restore the symmetry triggering kaon
condensation~\cite{BKR}. The transition is essentially triggered
by the ``relevant" mass term in the Lagrangian, so is S-wave.
There is a beautiful support for this dropping-mass scenario from
laboratory experiments~\cite{LLB}.

In strong interactions, strangeness has to be conserved and if
kaons were to condense, the pairs $K^\pm$ would have to do so. If
it were to happen at a low enough density, say, $\rho/\rho_0\sim 3$,
then condensation could occur only if the repulsive term that
counteracts the scalar attraction associated with the symmetry
breaking term, namely, the so-called vector coupling, gets
decoupled. In this case, both $K^\pm$ mesons become lighter and
eventually may condense. But if this were to occur at much higher
density than say $\sim 3\rho_0$, then kaons would be prevented
from condensing in heavy-ion process for the reason to be
mentioned below.

In the neutron-star matter, the kaon mass does not have to go all the
way to zero. In fact when the electron chemical potential that
increases with density -- for non-relativistic nucleons -- crosses
the dropping kaon mass, the electron will be converted by weak
interaction to the kaon
 \be
e^-\rightarrow K^- +\nu_e
 \ee
and the condensing of kaons is favored for energetic reasons. Here
$K^+$ will play no role. For $K^-$ the vector interaction is
attractive, so it speeds up the drop of the kaon mass.

The theoretical arguments that go into this treatment can be
reliable only if the condensation occurs at low enough density.
Were it to occur at much higher density, then there would be no
reason to believe that the approximation makes any sense. Indeed
the physics at higher density coming up in the bottom-up approach
will be obstructed by the inability to handle the short-distance
physics referred in nuclear physics to as short-range
correlations. For instance a straightforward embedding of
two-nucleon interactions at short distance into many-body systems
would predict that the phase transition will take place at a much
higher density than that relevant to neutron stars~\cite{pand}.
However this consideration is based on the Hartree approximation
in a potential model with a $KN$ potential that is not realistic
at high density~\cite{BR2000}. Nonetheless the issue of
short-range correlations is open.
 \vskip 0.5cm
\subsection{BR Scaling}
 \itt
Once kaons condense at some density $\rho_{Kc}$, going up further
in density up to (QCD) chiral restoration is like opening a
Pandora's box. As mentioned, we have no firm proof for {\it any}
transition with density since lattice calculations are not
feasible at present. There are a large number of different and
conflicting ideas published in the literature. I neither have
time nor wish to go into details here. It is fair to say that
nothing {\it additional} that is illuminating or startling is
expected to emerge in the physics of compact stars in this
regime, once the kaon condensation and related processes have
done what Chang-Hwan describes in his talk. In this murky
circumstance, I will take an option that allows me to go ahead
all the same. That is to heed what is said in Chinese fortune
cookies in North America: ``{\it There are only two ways of doing
things: One is the wrong way and the other is our way.}" I shall
therefore take ``our way" which is to invoke ``Brown-Rho (BR)
scaling"~\cite{BR91,BRPR}.

Consider symmetric nuclear matter. (Asymmetric matter can be
treated similarly.) At a density found in nuclear matter, one
finds that a description in which all degrees of freedom are {\it
quasi-particles} treated in the mean field works well provided
that those quasi-particles have masses and coupling constants
that respect chiral symmetry and run with
density~\cite{migdal-90,song}. In particular if one considers an
effective field theory with chiral symmetry with the nucleons,
pions, scalar meson and vector mesons as the appropriate degrees
of freedom, then with density, the masses run with density
as~\footnote{These are not {\it pole} masses. They can however be
related to physical masses given the appropriate corrections.}
 \be
\Phi (\rho) \approx \frac{f^\star_\pi (\rho )}{f_\pi} \approx
\frac{m^\star_\sigma (\rho )}{m_\sigma}\approx\frac{m_V^\star
(\rho )}{m_V} \approx \frac{M^\star (\rho )}{M}. \label{BRscaling}
 \ee
Here $M$ is the nucleon mass, and the subscripts $\sigma$ and $V$
stand respectively for the scalar $\sigma$ and the vector mesons $\rho,
\omega$. Since the pion decay constant can be taken as an order
parameter for chiral symmetry phases, the masses are all seen to
follow the order parameter in density. Thus when $f_\pi^\star
\rightarrow 0$ in the chiral limit as $\rho\rightarrow \rho_\chi$
(where $\rho_\chi$ is the as yet undetermined critical density
for chiral restoration), the masses approach zero in the same
way. In nature, the chiral limit is not reachable, so the masses
will not strictly go to zero but they will drop smoothly (in the
scenario of BR scaling) as density increases with interesting
physical consequences.

Following this BR scaling behavior all the way to the chiral
transition clearly sidesteps the Pandora's box alluded above. What
happens in this scenario is then that as one approaches the chiral
phase transition density, the relevant degrees of freedom will be
quasi-quarks (rather than quasi-nucleons), pions, scalar $\sigma$
(i.e., a chiral partner of the pion) etc. These will go {\it
smoothly} over to the other side of the phase boundary. This
smooth transition could also be witnessed when one comes down in
the top-down way from asymptotic density as I will discuss below.

The BR scaling as postulated in \cite{BR91} has support both from
theory~\cite{migdal-90,song} and from experiments, e.g., in
inelastic electron scattering from nuclei and in dilepton
production in heavy-ion collisions~\cite{BR2000}. It seems to
also account for the $J/\psi$ suppression observed at CERN that
is heralded to be the signal for the formation of quark-gluon
plasma~\cite{thomas}. This may be one more evidence for
``quark-hadron" continuity and/or ``Cheshire Cat phenomenon"
mentioned below.

 \vskip 0.5cm
 \section{Top-Down Approach}
 \subsection{Color superconductivity}
 \itt
When the density is asymptotically high, QCD in weak coupling is
applicable so one can compute things reliably. Because of
asymptotic freedom, the gauge coupling becomes weak and the
dynamics of the quarks is governed by one gluon exchange. In this
situation, diquarks are mostly likely to condense in the scalar
channel giving rise to color superconductivity. There have been
many papers written recently on this subject~\cite{csc}. I would
like to discuss a few aspects of the phenomenon that may be
relevant to the physics of compact stars.

When the quark chemical potential is so high that one can ignore
the u-, d- and s-quark masses, the situation becomes simple and
elegant. But as one goes down to the regime where the chiral phase
transition takes place, the situation gets very muddy and we have
no idea what's really happening. Even color superconductivity
scenarios differ depending upon how many flavors come into play.
At non-asymptotic chemical potential, other excitations than
diquark, such as quark-hole modes (e.g., Overhauser modes), can
be equally or even more important.  I shall therefore focus on
the ideal case of infinite chemical potential with three flavors
of quarks and hope for the best as we go down to the density
relevant to compact stars.

A left-handed quark is characterized by three colors (i.e., group
$SU(3)_c$) and three flavors (i.e., group $SU(3)_L$), so can be
denoted ${q_L^\alpha}_a$ with $\alpha$ standing for color and $a$
for flavor and similarly for the right-handed quark. It turns out
that it is most favorable to condense the diquark in the
color-flavor-locked (CFL) state as
 \be
\left<{q_L}_{i\alpha}^a(\vec p){q_L}_{j\beta}^b(-\vec p)\right>=
-\left<{q_R}_{i\alpha}^a(\vec p) {q_R}_{j\beta}^b(-\vec
p)\right>=\kappa(p_F)\epsilon_{\alpha\beta}
\epsilon^{abI}\epsilon_{ijI}
 \ee where $i, j$ are the spin indices.
This means that both color and chiral symmetries are
spontaneously broken by the condensate as~\footnote{The $U(1)_V$
corresponding to the fermion number is also broken but the
$U(1)_A$ symmetry is restored at high density. There is also a
$Z_3$ invariance that remains but we need not be concerned with
it here.}
 \be SU(3)_c\times SU(3)_L\times
SU(3)_R\rightarrow SU(3)_{c+L+R}.
 \ee
 As a consequence, seventeen Goldstone bosons
are excited, i.e., nine scalars and eight pseudoscalars. Eight
scalars get eaten up by the gluons which become massive and the
remaining scalar gets condensed to give rise to superfluidity.
The eight pseudoscalars constitute the lowest modes remaining in
the system that pick up a mass proportional to the current quark
mass that goes to zero as $\mu\rightarrow\infty$.

The most remarkable conjecture made by Sch\"afer and
Wilczek~\cite{cont} posits that there is one-to-one correspondence
between the spectrum of excitations in the zero-density regime
and that of excitations in the CFL phase. Apart from the 8
Goldstone pseudoscalars that are present in both regimes, the
eight massive gluons possess the same quantum numbers as the
octet vector mesons in free space and the 8 color-flavor-locked
quarks (which may be simply understood as quark solitons or
qualitons~\cite{hrz}) can be mapped to the octet baryons of zero
density. The conjecture is that the infinite-density spectrum is
{\it continuously connected} to the zero-density spectrum,
implying ``quark-hadron continuity" or in the more pictorial term,
``Cheshire Cat phenomenon"~\cite{NRZ}. This aspect shows that
going toward high density beyond $\rho_\chi$ may reveal a richer
and quite different variety than going to high temperature. There
is however an obstacle to this intriguing possibility: there may
be a series of phase changes as one goes down in density from
infinite to the regime relevant to compact stars. First of all,
even in color superconductivity (CSC), scenarios differ depending
upon the scale of the strange quark mass relative to the chemical
potential~\cite{csc}. Not less importantly, there can be other
phases than color superconductivity such as Overhauser
effect~\cite{PRWZ}, ISB~\cite{ISB} and others~\cite{KR} that may
intervene before going into the Goldstone phase of lower density.
Despite the plethora of other possibilities, it is remarkable --
and perhaps meaningful -- that both pion
condensation~\cite{sonpion} and kaon
condensation~\cite{schaeferkaon} could occur in the high density
regime. If confirmed, this could be another support for the
notion of quark-hadron continuity.
 \subsection{Astrophysical implications}
  \itt
 The mere existence of pions and kaons in the CFL phase is a highly
intriguing possibility one would like to explore. Although
quarks/baryons are gapped in the CFL phase with a gap $\Delta\sim
10-100$ MeV, there can still be light fermions in some situations.
These degrees of freedom could play an important role in
astrophysical compact objects. Since it does not look feasible to
make firm predictions for both terrestrial and astrophysical
observables, one would like to turn the process around and {\it
learn} from astrophysical observations about the state of matter
in which such exotic phenomena take place.

There have been a series of papers addressing the role of color
superconductivity (CSC) in the cooling of neutron
stars~\cite{cooling}. The situation is quite complex and a
complete analysis is probably not yet possible. The complexity
comes from the fact that we do not really know what should be
taken into account in the cooling since there are meson
condensations and other processes that could intervene at the
same time. Even within the framework of CSC, there can be several
competing factors to consider. First there is the effect on the
electro-weak interaction proper. A simple calculation shows that
inelastic neutrino processes (i.e., neutrino production) can be
enhanced in the CFL phase~\cite{HLNR}. Next, the produced
neutrinos may have longer mean-free path in the CFL phase than in
normal phase~\cite{carter}, thereby increasing the emissivity. On
the other hand, the gap $\Delta$ can be large so that there will
be a strong suppression from the Boltzmann factor
$exp(-\Delta/T)$. All these competing factors would have to be
carefully accounted for before one can come to any estimate of
the cooling rate.  A lot more work and help from observations will
be needed for a clearer picture. We heard in this conference
about a very interesting data coming from a point source in
Cassiopeia A which could be a valuable clue to the
issue~\cite{nomoto}.

Another interesting effect is on magnetic fields in compact
stars~\cite{alfordmag,blaschkemag}. Since vector symmetry is
spontaneously broken (actually in all scenarios of color
superconductivity (CSC)), both color and ordinary electromagnetism
are spontaneously broken. However although electrically charged
diquarks condense as do electron pairs in ordinary
superconductivity, they do not produce the familiar Meissner
phenomenon of ordinary superconductivity. The reason is that
there is an unbroken $U(1)$ symmetry with a massless gauge boson.
What happens is that as long as the color gauge coupling $g_c$ is
much bigger than the electromagnetic coupling $e$, magnetic
fields can penetrate into the system largely un-expelled from the
CSC phase. As a consequence, if rotating compact stars are in the
CSC phase, the magnetic field will be stable against decay on time
scales longer than, say, $10^7$ years set by
observation~\cite{makashima}. In \cite{alfordmag}, it is even
suggested that if an isolated pulsar were observed to lose its
magnetic field as it spins down, it would rule out the presence
of the CSC (or even quark) phase in the pulsar. Although quite
interesting, this negative evidence is not too exciting. It would
be a lot more exciting if the CFL or other CSC phases were
associated with a positive consequence. Again, given the various
other possibilities in the density regime involved, the situation
may be different from this simplified consideration. It remains
an open issue.
\section{Summary}
\itt
 The approach to the density regime relevant to compact stars is
 made in both ways: ``bottom-up" and ``top-down." The two
 extremes, namely, bottom and top, are fairly well described, the
 former based on effective field theory of QCD combined with
 experimental data and the latter based on weak-coupling theory of
 QCD. The two approaches leave a rather wide gap in between
 where they have no overlap and it is in this range that
 interesting phenomena seem to take place in dense matter in compact
 stars. A bridge is suggested in terms of BR scaling coming from
 bottom-up. Combined with what we learn from the top-down approach
 and from astrophysical observations, it would be possible to
 establish the intriguing ``quark-hadron continuity" alias
 ``Cheshire Cat." The main aim of this talk was to stress that the
 compact stars are the only source for physics at high density,
 the domain that cannot be accessed in terrestrial laboratories.

{\subsection*{Acknowledgments}}

 It is a pleasure to acknowledge the generous help and hospitality
 of Korea Institute for Advanced Study for my participation in
 co-organizing, and for attending,  this first KIAS workshop on
 astrophysics and astro-hadron physics. This paper was prepared
 while I was visiting Korea Institute for Advanced Study.

\end{document}